\documentclass{article}
\usepackage{graphicx}
\usepackage{amsmath, amssymb}

 \usepackage[dblblindworkshop, final]{neurips_2025}

\workshoptitle{Machine Learning and the Physical Sciences}

\usepackage[utf8]{inputenc} 
\usepackage[T1]{fontenc}    
\usepackage{hyperref}       
\usepackage{url}            
\usepackage{booktabs}       
\usepackage{amsfonts}       
\usepackage{nicefrac}       
\usepackage{microtype}      
\usepackage{xcolor}         

\title{Multi-modal Foundation Model \\ for Cosmological Simulation Data}

\author{%
Bin Xia\thanks{Corresponding author. Work completed during an internship at Argonne National Lab.} \\
Center for Relativistic Astrophysics, School of Physics \\
Georgia Institute of Technology \\
Atlanta, GA 30332 \\
\texttt{xiabin@gatech.edu} 
\And
Nesar Ramachandra\thanks{These authors contributed equally.} \\
Computational Science Division\\
Argonne National Laboratory\\
Lemont, IL 60439 \\
\texttt{nramachandra@anl.gov} 
\And
Azton I. Wells\footnotemark[2] \\
Computational Science Division\\
Argonne National Laboratory\\
Lemont, IL 60439 \\
\texttt{awells@anl.gov} 
\And
Salman Habib \\
Computational Science Division\\
Argonne National Laboratory\\
Lemont, IL 60439 \\
\texttt{habib@anl.gov} 
\And
John Wise \\
Center for Relativistic Astrophysics, School of Physics \\
Georgia Institute of Technology \\
Atlanta, GA 30332, USA \\
\texttt{jwise@physics.gatech.edu} 
}

\begin{document}

\maketitle

\begin{abstract}
We present a multi-modal foundation model for astrophysical galaxy data, designed to map between simulation- and observation-based galactic features. Our encoder-only transformer flexibly ingests scalar quantities (e.g., redshifts, galaxy masses) and vectors (e.g., star formation histories, spectra), supporting multi-task training that includes within-modality reconstruction and cross-modality prediction. With a dynamic masking strategy, the model can query arbitrary galaxy properties from partial inputs—including predicting spectra from redshift and mass, or estimating photometric redshifts from broadband magnitudes—while also recovering missing segments within a modality. Trained on 185,000 simulated galaxies from a gigaparsec-scale Cosmology simulation, the model yields a 50\% improvement in redshift estimation when combining LSST and SPHEREx photometry over LSST photometry alone, and a 63\% improvement in stellar mass inference when combining late-time SFH with LSST photometry over early-time SFH with LSST photometry. The model demonstrates strong generalization across multi-modal tasks and lays the groundwork for future integration of higher-dimensional and structured data such as images, merger trees, and 3D fields. This approach provides a unified framework for connecting simulations and observations, advancing the development of generalizable astrophysical foundation models. 
\end{abstract}

\section{Introduction}
Understanding the connection between cosmological simulations and astronomical observations is a central challenge in modern astrophysics \cite{vogelsberger2020cosmological, 2017ARA&A..55...59N, 2015ARA&A..53...51S}. Large-scale simulations, such as the Last Journey project created using the Hardware/Hybrid Accelerated Cosmology Code (HACC) \cite{osti_1249547, 2021ApJS..252...19H}, provide detailed predictions of dark matter structures and their evolution. Galaxies, their formation histories, and synthetic observables are subsequently derived from these simulation outputs through post-processing heuristic models of galaxy-halo connection and stellar population synthesis \cite{korytov2019cosmodc2, 2023MNRAS.518..562A}. Modern hydrodynamical simulations \cite{pakmor2023millenniumtng, frontiere2023simulating} directly simulate galaxies using gas dynamics and subgrid models, in addition to gravity, and require fewer post-processing steps to obtain observable quantities. Meanwhile, wide-field surveys from Rubin Telescope \cite{le2024rubin} and SPHEREx \cite{2020SPIE11443E..0IC} will deliver high-dimensional observational data across millions of galaxies. Bridging these two domains—simulation-derived physical quantities and observation-driven measurements—is essential for cosmological inference, yet remains difficult due to their heterogeneous representations and incomplete overlap.

Traditional machine learning approaches in cosmology often focus on narrow, single-modality tasks, such as photometric redshift estimation \cite{newman2022photometric, d2018photometric} or classifying strong lenses \cite{shajib2024strong}. These models rely on hand-crafted datasets and are not extendable across diverse data sources and representations. 
In contrast, the success of foundation models in other scientific and engineering domains suggests a new paradigm: transformer-based architectures trained on large \cite{jumper2021highly, radford2021learning, schutt2017schnet}, heterogeneous, multi-modal datasets can capture general-purpose representations that support flexible downstream tasks. For instance, Gloeckler et al. \cite{2024arXiv240409636G} show that transformer-based probabilistic diffusion models can perform flexible amortized Bayesian inference on simulation-based models, handling missing or unstructured data across diverse tasks.
In astrophysics, efforts such as AstroCLIP \cite{parker2024astroclip} have demonstrated the potential of contrastive multi-modal pretraining for aligning images and catalogs. A foundation model for stars \cite{Leung_2023} has demonstrated that transformer-based approaches can effectively unify the modalities of stellar data. Existing efforts, such as OmniJet-$\alpha$ \cite{birk2024omnijet}, target specific domains in particle physics, while large-scale multimodal datasets, like the Multi-Modal Universe (MMU) \cite{2024arXiv241202527T}, underscore the need for truly general-purpose models. These advances indicate that astronomy, with its rich and heterogeneous data landscape, is well-positioned to benefit from such a foundation model approach.

Motivated by these developments, we present a multi-modal foundation model for cosmological simulation data.
We adopt an encoder-only transformer, similar to BERT’s encoder \cite{devlin2019bert}, that jointly encodes scalar and vector modalities, including redshift, halo mass, stellar mass, star formation histories (SFHs), photometric magnitudes, and spectral energy distributions (SEDs). To enable cross-modality and within-modality reconstruction, we develop a novel dynamic masking strategy within this framework. By training on a subset of galaxies derived from the Last Journey simulation, we demonstrate that the model learns a coherent latent representation supporting flexible inference tasks such as photometric redshift estimation and missing data reconstruction.

\section{Model Architecture and Methodology}

We introduce \textbf{MOSAIC} (short for Multi-modal Observation-Simulation Integration for Cosmology), 
a model designed to enable translation and joint understanding of simulation-only quantities and synthetic observables. Our approach leverages an encoder-only transformer, inspired by BERT’s design \cite{devlin2019bert}, to learn shared representations across diverse scalar and vector modalities derived from the massive-volume HACC cosmological simulations \cite{osti_1249547, 2021ApJS..252...19H}. These modalities encompass both simulation-only quantities (e.g., dark matter halo masses) and have been post-processed to include synthetic observations (e.g., photometric magnitudes and SEDs, allowing the model to bridge the gap between simulation and observation through a unified latent space.

\textbf{Data Types and Normalization.}
Our dataset comprises 185,247 training samples and 20,583 test samples from an extragalactic catalog generated from the Last Journey simulation \cite{2021ApJS..252...19H}. While this gravity-only simulation includes only dark matter, we assign galaxies, their formation histories, associated spectra, and broad magnitudes onto the SMACC dark matter cores using a combination of galaxy–halo connection models, stellar population synthesis, and heuristic fits to observational data \cite{2021ApJS..252...19H, 2021ApJ...913..109S}. Each galaxy sample in the resulting extragalactic catalog comprises a mixture of scalar and vector modalities, encompassing a broad range of galaxy types, environments, and formation histories. However, we note that the distributions of the galactic properties are not representative of any single survey. This extragalactic dataset is a randomly chosen subset used to train the foundation model and evaluate the model's ability to perform multi-modal reconstruction and prediction. Realistic extragalactic datasets like this have been successfully deployed on observational targets with pre-defined mapping, like photometric redshift estimations \cite{ramachandra2022machine}, or strong lensing parameter estimation \cite{lanusse2018cmu}, but have not been fully utilized in a `task-agnostic' setting of a foundation model.  

We categorize the synthetic catalog entries into scalar and vector modalities. Scalar inputs (zero-dimensional), including \textit{redshift}, \textit{halo mass}, and \textit{stellar mass}, are standardized by subtracting the mean and dividing by the standard deviation computed over the entire training set for each scalar. Vector inputs (one-dimensional) include photometric magnitudes (AB-magnitudes in 6-band LSST (ugrizY) and 102 near-infrared colors from SPHEREx), star formation histories (117 cosmic-time bins, in log-scale units of $\log_{10}{\rm M_\odot/yr}$), and rest-frame spectral energy distributions (921 wavelength bins, in units of $\log_{10}\rm Jy$).
Each vector modality is normalized by subtracting its global mean and then dividing by its standard deviation, which is computed over the entire training set.

\begin{figure}[htbp]
    \centering
    \includegraphics[width=\textwidth]{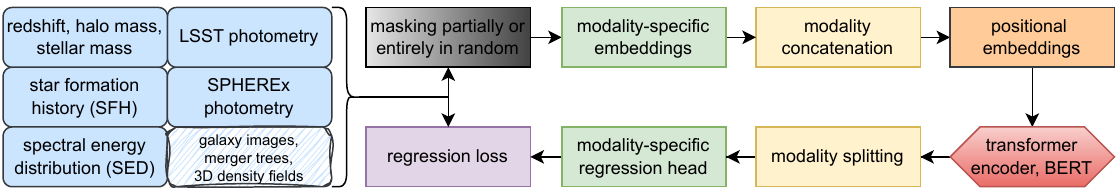}
    \caption{Schematic of MOSAIC architecture. Each modality is masked and projected into modality-specific embeddings, concatenated along the sequence dimension, combined with positional embeddings, and processed by a Transformer encoder. The sequence is split into modality-specific segments and fed into dedicated regression heads to produce predictions and compute regression loss. 
    }
    \label{fig:architecture_MMFM}
\end{figure}

\textbf{Model Architecture.}
As illustrated in Fig.~\ref{fig:architecture_MMFM}, MOSAIC handles scalar and vector modalities using an encoder-only Transformer backbone, similar to BERT \cite{devlin2019bert}. To enable flexible inference from incomplete or heterogeneous data, each modality is first randomly masked, either fully or in contiguous blocks—simulating stretches of missing data—following the hierarchical scheme described in Sec.~\ref{sec:masking}.  Physically, masking corresponds to simulating missing observations or unmeasured properties for a galaxy. In practice, masked entries are set to zero in our experiments.

The resulting masked and unmasked inputs are projected into modality-specific embeddings (see Sec.~\ref{sec:input_representation}), which map each physical quantity into a continuous latent space. Physically, these embeddings capture the relative magnitudes, patterns, or shapes of the underlying scalar and vector quantities, allowing the model to learn relationships between different galaxy properties. The embeddings from all modalities are then concatenated along the sequence dimension, combined with shared positional embeddings via element-wise addition, and processed by the Transformer encoder. After encoding, the sequence is split back into modality-specific segments, each of which is passed through its dedicated regression head. The model is trained to perform both in-modality reconstruction (predicting missing values within a modality, e.g., SED reconstruction when only visible spectra are available) and cross-modality prediction (predicting one modality from another, e.g., estimating photometric magnitudes from halo mass and SFH). The predictions are compared to the corresponding ground truth to compute regression loss, as detailed in Sec.~\ref{sec:training_loss}.

\section{Results}

\begin{figure}[t]
    \centering
    \includegraphics[width=\linewidth]{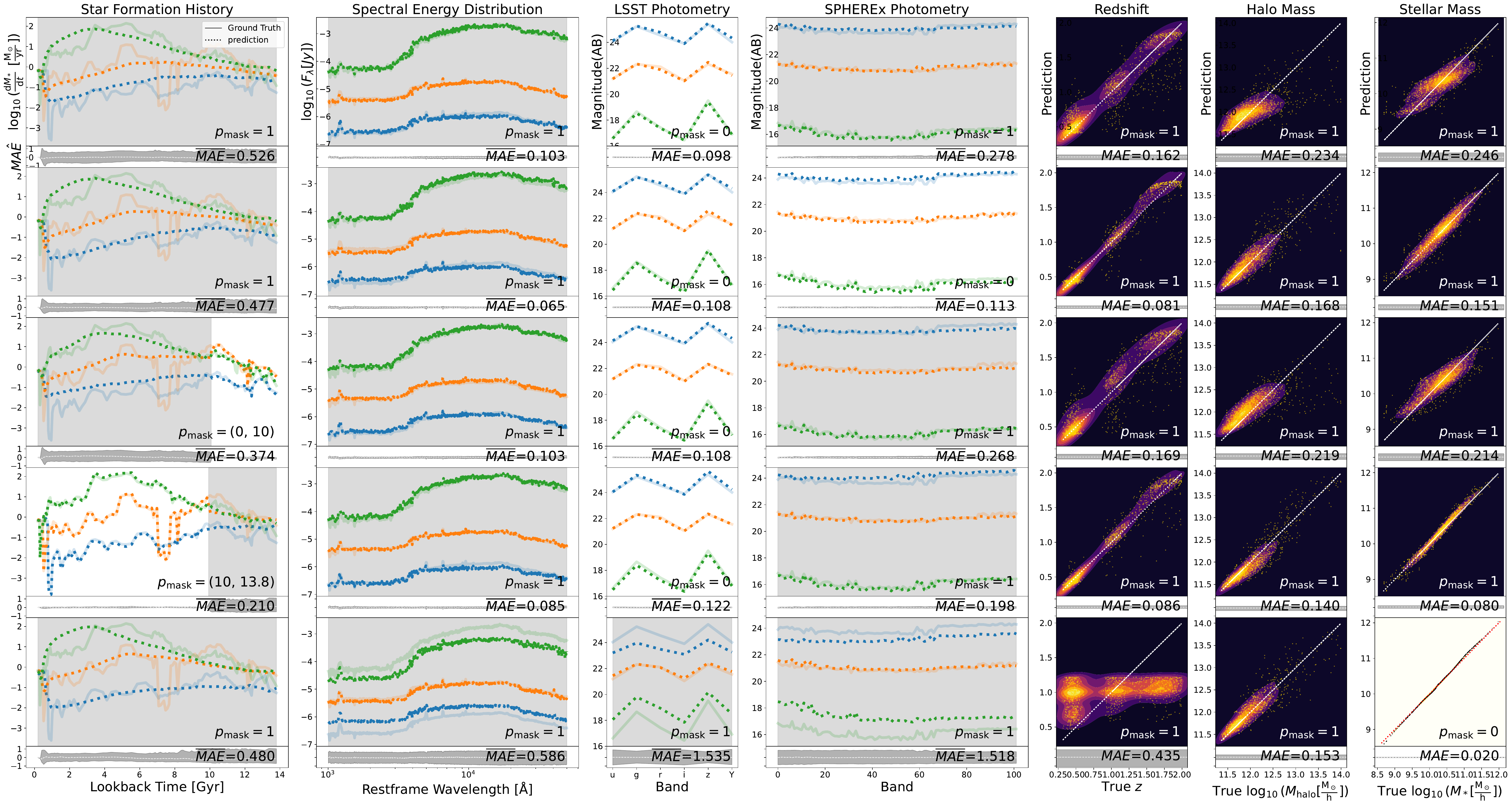}
    \caption{
    Illustration of multi-modal prediction tasks and input masking configurations. 
    Each row corresponds to a different input combination: (1) LSST magnitudes only, (2) LSST magnitudes + SPHEREx colors, (3) LSST magnitudes + early SFH (0--10 Gyr masked), (4) LSST magnitudes + late SFH (10--13.8 Gyr masked), (5) stellar mass only. 
    White areas indicate provided data, and shaded areas indicate masked portions. 
    For vector modalities, each subplot shows three samples with light solid lines for ground truth and dark dotted lines for predictions. 
    For scalar modalities, each subplot displays scatter points for 1000 samples, with points near the diagonal indicating accurate predictions. 
    Shaded bands indicate the 16–84\% range ($\approx 1\sigma$) of the normalized mean absolute error (MA$\hat{\text{E}}$) computed over 1000 samples; unnormalized mean absolute error ($\overline{\text{MAE}}$ for vectors and MAE for scalars) is annotated for reference. 
    }
    \label{fig:results_schematic}
\end{figure}

We evaluate MOSAIC on five representative input configurations. For each configuration, the model receives partial observations and predicts all target astrophysical quantities. This setup illustrates MOSAIC's flexibility in handling diverse combinations of available catalog inputs and performing both cross-modality prediction and partial reconstruction within a modality. According to Fig.~\ref{fig:results_schematic}, key prediction performance, measured by the mean absolute error (MAE), is summarized in Table~\ref{tab:prediction_mae_compact}, with smaller values indicating better performance. Some entries are omitted, as they correspond to input-target combinations that are less informative for this illustrative problem.

\begin{table}[ht!]
\centering
\caption{Mean Absolute Error (MAE) for predictions of redshift, SFH (averaged over time), and stellar masses with different input modalities. Smaller is better.}
\label{tab:prediction_mae_compact}
\begin{tabular}{l c c c}
\hline
\textbf{Input Modalities} & \textbf{Redshift} & \textbf{SFH} & \textbf{Stellar mass} \\
\hline
LSST magnitudes only & 0.162 & 0.526 & - \\
LSST magnitudes + SPHEREx colors & 0.081 & 0.477 & - \\
LSST magnitudes + early SFH & 0.169 & - & 0.214 \\
LSST magnitudes + late SFH & 0.086 & - & 0.080 \\
Stellar mass only & 0.435 & 0.480 & - \\
\hline
\end{tabular}
\end{table}

These results indicate several key insights. For stellar mass prediction, late SFH combined with LSST photometry provides the strongest constraint, whereas for redshift estimation, combined photometry (LSST + SPHEREx) or LSST with late SFH achieves the highest accuracy. LSST alone yields moderate performance, demonstrating that multi-modal inputs substantially improve predictive capability. 
Interestingly, using stellar mass alone as input, while insufficient to accurately predict SED, photometry, or redshift, still allows the model to recover the overall trend of the SFH and provides reasonable estimates of halo mass.
Certain tasks, such as SFH prediction from LSST only, remain challenging, reflecting the necessity of complementary modalities to fully constrain complex vector quantities. 
These experiments demonstrate MOSAIC’s ability to perform multi-task prediction from heterogeneous inputs while highlighting which input combinations are most informative for different astrophysical properties.
Scatter plots show some clumping patterns, which arise from the non-uniform sampling of the training and test sets: oversampled parameter regions lead to denser clusters along the $x$-axis, while the model’s predictions in high-uncertainty regions concentrate around the most likely outcomes seen in the training set, producing vertical clustering.

\begin{figure}[t]
    \centering
    \includegraphics[width=\linewidth]{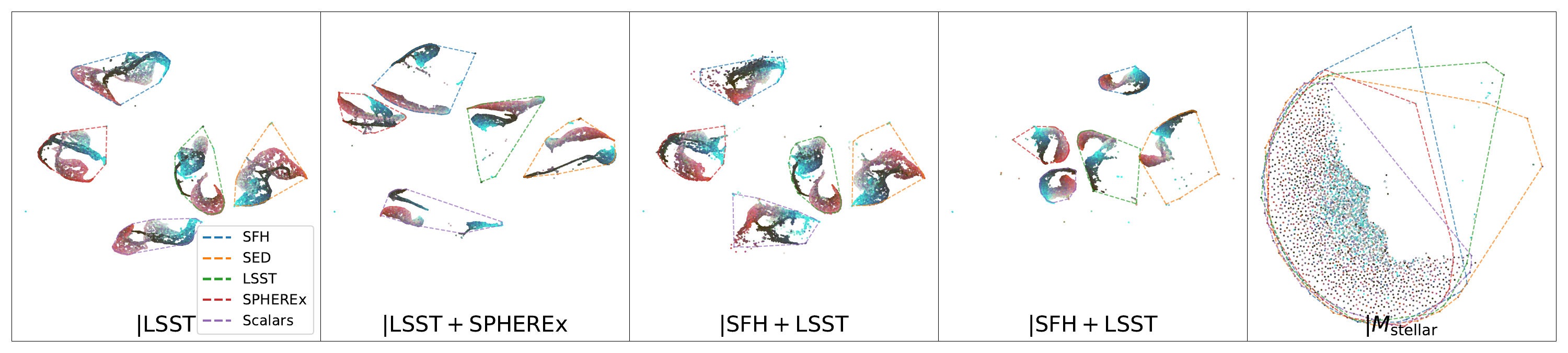}
    \caption{
    UMAP visualization of the last hidden state embeddings for five input configurations corresponding to Fig.~\ref{fig:results_schematic}. 
    Each subplot shows embeddings of five modalities (SFH, SED, SPHEREx, LSST, Scalars) for 10000 galaxies. 
    Convex hulls are drawn around points between the 0.1 and 99.9 percentiles to reduce outlier effects, with a distinct color for each modality. 
    Point colors encode normalized scalar ground-truth values (redshift, halo mass, stellar mass). 
    For the first four configurations, embeddings form well-separated clusters corresponding to each modality, and within each cluster, points with similar physical properties (similar colors) are located close to each other, forming smooth color gradients. 
    This indicates that the latent space captures astrophysical correlations even when input scalars are masked. 
    For the fifth configuration, where only stellar mass is provided, clusters overlap and color continuity is absent, consistent with weaker predictive performance.
    The axes correspond to the two UMAP embedding dimensions, which are abstract and unitless.
    }
    \label{fig:latent_umap}
\end{figure}

To further investigate the physical insights captured by the model, we visualize the last hidden state embeddings using UMAP for the five input configurations in Fig.~\ref{fig:latent_umap}.
Within each subplot, embeddings from the five modalities (SFH, SED, SPHEREx, LSST, Scalars) are shown for 10000 samples, enclosed by convex hulls that exclude extreme outliers (points outside the 0.1--99.9\% range in $x$ and $y$). 
For the first four input configurations, embeddings of different modalities form well-separated clusters, indicating that the model can effectively distinguish and reconstruct distinct modalities. Furthermore, within each cluster, points are colored by their normalized ground-truth scalar properties (redshift, halo mass, stellar mass), and galaxies with similar physical properties lie closer together, producing smooth color gradients. This demonstrates that MOSAIC captures meaningful astrophysical correlations in its latent space, even when those scalar quantities are fully masked at the input stage. By contrast, when stellar mass alone is provided (fifth configuration), embeddings lack clear cluster separation,
consistent with the weaker predictive performance shown in Fig~\ref{fig:results_schematic} and Table~\ref{tab:prediction_mae_compact}.

\section{Discussion and Conclusion} 
Multiwavelength astronomy has proven highly effective, as different surveys probe distinct physical processes \cite{meszaros2019multi, troxel2023joint}. Likewise, simulations provide correlated quantities—such as formation histories and density environments—that, if brought into a unified framework, offer significant opportunities for new physical insights.
MOSAIC can flexibly predict galaxy properties from partial inputs: it can perform photometric redshift and stellar mass estimation, reconstruct SEDs and SFHs, and translate between photometric and spectroscopic measurements from different telescopes. For example, combining LSST and SPHEREx photometry yields an MAE of $0.081$ for redshift estimation, while stellar mass can be inferred from late-time SFH combined with LSST photometry with an MAE of $0.080$, substantially outperforming early-time SFH (MAE = $0.214$). These results demonstrate the complementary roles of photometric and spectroscopic inputs, highlighting the model’s ability to integrate heterogeneous data for robust cross-modality prediction. 

Predictive accuracy varies across properties, reflecting underlying astrophysical correlations. Quantities with direct observational imprints, such as redshift from photometric colors or stellar mass from late-time SFH, are more strongly constrained and yield higher accuracy. In contrast, reconstructing full star formation histories from photometry alone remains challenging, since cumulative SED features do not uniquely encode detailed temporal evolution. This explains why some tasks exhibit stronger performance than others and highlights the complementary role of multi-modal inputs. 

Compared with existing foundation models in astronomy, which are trained solely on observational data \cite{parker2024astroclip, Leung_2023}, MOSAIC is the first framework designed for large-scale cosmological simulation datasets, aiming to bridge simulations and observations. A key strength in our approach is the masking-based training strategy, which enables learning from incomplete and heterogeneous data. 
We note that, since our current training relies on simulated samples, predictive performance may partly reflect simulator-dependent biases; future work will incorporate observational datasets when scaling to larger data volumes to mitigate such biases. 
The clumping patterns observed in Fig.~\ref{fig:results_schematic} exemplify such sampling-driven biases, 
highlighting the need for careful treatment of training distributions.
Moreover, analysis of the latent space (Fig.~\ref{fig:latent_umap}) suggests that MOSAIC embeddings capture meaningful astrophysical correlations, with similar galaxies mapped nearby even when input scalars are masked, providing additional interpretability beyond raw predictive accuracy.

While the current implementation focuses on point predictions, the learned multi-modal latent space could naturally support more complex generative tasks. For instance, adding a probabilistic decoder on top of the latent embeddings would allow joint generative modeling of multiple galaxy properties, enabling uncertainty quantification and conditional sampling. Approaches such as variational autoencoders, normalizing flows, or diffusion models could be integrated into the framework to produce full posterior distributions rather than single-point estimates, extending the model’s applicability to simulation-based inference, synthetic data generation, or probabilistic survey planning.

There remains scope for further enhancements: extending to higher-dimensional modalities such as galaxy images, 3D density fields, or merger trees, requiring modality-specific encoders (e.g., convolutional or graph-based) for spatial data before integration into the shared transformer latent space;
handling domain shifts when applying to real surveys; exploring pretraining strategies; 
and evaluating downstream cosmological analyses such as weak lensing and galaxy clustering. 
In the near term, MOSAIC could be applied directly to observational datasets to infer missing galaxy properties or to simulations to generate synthetic observables, offering a versatile tool for survey planning, analysis, and astrophysical interpretation.

\begin{ack}
This material is based upon work supported by Laboratory Directed Research and Development (LDRD) funding from Argonne National Laboratory, provided by the Director, Office of Science, of the U.S. Department of Energy under Contract No. DEAC02-06CH11357. The training is conducted on Swing, a GPU system at the Laboratory Computing Resource Center (LCRC) of Argonne National Laboratory, and utilizes the resources of the National Energy Research Scientific Computing Center (NERSC), a Department of Energy User Facility, with an NERSC award, GenAI@NERSC.  
\end{ack}



\small

\bibliographystyle{unsrt} 
\bibliography{references}  


\appendix
\section{Masking Strategy}
\label{sec:masking}

To support both in-modality reconstruction and cross-modality prediction, we employ a hierarchical masking scheme applied at data loading time. This design ensures balanced training across different modality combinations while also simulating realistic scenarios of incomplete data.

\paragraph{Implementation Details of Masking.}
Masked entries are set to a fixed value, $\texttt{mask\_token} = 0$, corresponding to the mean of each modality after normalization. This value lies within the physical range and has been empirically found to yield the best performance. Both masked and unmasked inputs are passed through the model during training, allowing it to leverage available information while reconstructing missing values. This approach stabilizes training and improves generalization across diverse cross-modal and in-modality prediction tasks.

\paragraph{Global Modality Masking.}
Each modality, scalar or vector, is independently masked entirely with probability $0.5$. 
Consequently, all $2^N$ possible modality subsets (for $N$ modalities) occur with equal probability $2^{-N}$, 
ensuring balanced exposure to every cross-modal configuration during training.

\paragraph{Masking for Scalar Modalities.}
All scalar inputs (e.g., redshift, halo mass, stellar mass) are concatenated into a single vector, denoted as \textit{scalars}. 
Each element within this vector is independently masked with probability $0.5$. 
Unlike vector modalities, scalars do not undergo partial token masking; masking corresponds simply to replacing the value with $\texttt{mask\_token}$.

\paragraph{Local Token Masking for Vector Modalities.}
For vector modalities (e.g., SED or SFH) that are not fully masked, we apply \emph{partial token masking}. 
A masking ratio $p=0.2$ defines the total number of tokens to mask:
\[
N_{\text{mask}} = \lceil p \cdot d \rceil,
\]
where $d$ is the sequence length. Masking is performed iteratively by sampling spans of random lengths and positions until the budget $N_{\text{mask}}$ is exhausted. By sampling spans of varying lengths, the masked tokens include both isolated points and contiguous segments, encouraging the model to learn local interpolation as well as global distributional patterns. Since the masking ratio is relatively small, overlaps between spans are allowed.

\section{Input Embedding and Representation}
\label{sec:input_representation}

All inputs are organized into modality-specific token sequences. 
In particular, all scalar quantities (e.g., redshift, halo mass, stellar mass) are concatenated into a single modality denoted as \textit{scalars}, so that they can be treated uniformly as a sequence of tokens. 
Vector modalities (e.g., SED, SFH) naturally form token sequences according to their sampled dimensions.

Each token, whether from \textit{scalars} or a vector modality, is mapped into the hidden space via a modality-specific linear projection.
For a given modality $M$, let $x_{b,\ell}^{(M)}$ denote the value of token $\ell$ in batch $b$, and let $L_M$ be the sequence length of modality $M$. 
Each token is mapped into the hidden space via a modality-specific linear projection:
\[
\mathbf{h}_{b,\ell}^{(M)} = W_M x_{b,\ell}^{(M)} + b_M,
\qquad W_M \in \mathbb{R}^{d_{\text{hidden}}\times 1},\; b_M \in \mathbb{R}^{d_{\text{hidden}}},
\]
yielding the embedded representation $\mathbf{H}^{(M)} \in \mathbb{R}^{B \times L_M \times d_{\text{hidden}}}$ for modality $M$.

This design avoids discrete tokenization, which is natural for text but suboptimal for continuous scientific data, while preserving numerical continuity and allowing each modality to learn a flexible embedding in a shared representation space. 
All embedded modalities are then concatenated, augmented with positional encodings, and processed by a shared Transformer encoder.

\section{Training Strategy and Loss Function}
\label{sec:training_loss}
We adopt masked regression as the primary training objective. 
For each modality $M$ with sequence length $L_M$, let $m_{b,\ell}^{(M)} \in \{0,1\}$ denote the mask indicator ($1$ if masked, $0$ otherwise). 
The model receives
\[
\tilde{x}_{b,\ell}^{(M)} =
\begin{cases}
\texttt{mask\_token}, & m_{b,\ell}^{(M)} = 1,\\[1ex]
x_{b,\ell}^{(M)}, & m_{b,\ell}^{(M)} = 0,
\end{cases}
\]
and produces predictions $\hat{x}_{b,\ell}^{(M)}$. In practice, we set $\text{mask\_token} = 0$, which corresponds to the global mean of normalized data, as we found this choice performs best across modalities. The per-token loss is
\[
\mathcal{L}_{b,\ell}^{(M)} = \bigl(\hat{x}_{b,\ell}^{(M)} - x_{b,\ell}^{(M)}\bigr)^2.
\]

The loss for modality $M$ is the mean over all tokens in the batch:
\[
\bar{\mathcal{L}}^{(M)} = \frac{1}{B L_M} \sum_{b=1}^{B} \sum_{\ell=1}^{L_M} \mathcal{L}_{b,\ell}^{(M)}.
\]

Finally, the overall training objective averages equally over all $K$ modalities:
\[
\mathcal{L} = \frac{1}{K} \sum_{M=1}^{K} \bar{\mathcal{L}}^{(M)}.
\]

\noindent\textbf{Why include unmasked tokens?}  
While the loss on masked tokens drives imputation and cross-modal reasoning, 
we also include reconstruction loss on unmasked tokens. 
This acts as a form of curriculum: initially, the model can reliably learn the trivial identity mapping, which stabilizes optimization and anchors the representation space. 
Once this foundation is established, the model progressively learns to leverage contextual and cross-modal information to recover masked values. 
In this way, unmasked-token supervision accelerates convergence, regularizes training, and improves overall robustness.  

Moreover, all tokens within a modality are treated with equal weight, regardless of whether they are involved in identity reconstruction, imputation, or cross-modality prediction. 
Similarly, losses from different modalities are averaged with equal weight, independent of their token counts. 
This design avoids bias toward particular tasks or modalities, ensuring a balanced and unbiased training signal that promotes generalizable representations.


\end{document}